\documentclass{article}

\usepackage{enumitem}
\usepackage{pifont}
\usepackage[dvipsnames]{xcolor} % 用于颜色
\usepackage{subfig}
\usepackage{float}
\usepackage{overpic}
\usepackage{multirow}
\usepackage{graphicx}
\usepackage{svg} 
\usepackage{colortbl}
\usepackage{amsmath}
\usepackage{wrapfig}
\usepackage{makecell}
\usepackage{listings}
\usepackage[preprint]{neurips_2025}

\usepackage[utf8]{inputenc} % allow utf-8 input
\usepackage[T1]{fontenc}    % use 8-bit T1 fonts
\usepackage{hyperref}       % hyperlinks
\usepackage{url}            % simple URL typesetting
\usepackage{booktabs}       % professional-quality tables
\usepackage{amsfonts}       % blackboard math symbols
\usepackage{nicefrac}       % compact symbols for 1/2, etc.
\usepackage{microtype}      % microtypography
\usepackage{xcolor}         % colors

\title{RAG-IGBench: Innovative Evaluation for RAG-based Interleaved Generation in Open-domain Question Answering}

\author{
    \textbf{Rongyang Zhang}$^{1,2}$,
    \textbf{Yuqing Huang}$^{1,2}$,
    \textbf{Chengqiang Lu}$^{2}$,
    \textbf{Qimeng Wang}$^{2}$,
    \textbf{Yao Gao}$^{2}$, \\
    \textbf{Yi Wu}$^{2}$,
    \textbf{Yao Hu}$^{2}$,
    \textbf{Yin Xu}$^{1}$,
    \textbf{Wei Wang}$^{3}$,
    \textbf{Hao Wang}$^{1}$,
    \textbf{Enhong Chen}$^{1}$ \\ \\
    $^1$State Key Laboratory of Cognitive Intelligence, University of Science and Technology of China \\
    $^2$Xiaohongshu Inc. \quad 
    $^{3}$Xi’an Jiaotong University \\
}

\newcommand{\benchmark}{RAG-IGBench }
\begin{document}

\maketitle

\begin{abstract}
  In real-world scenarios, providing user queries with visually enhanced responses can considerably benefit understanding and memory, underscoring the great value of interleaved image-text generation. Despite recent progress, like the visual autoregressive model that unifies text and image processing in a single transformer architecture, generating high-quality interleaved content remains challenging. Moreover, evaluations of these interleaved sequences largely remain underexplored, with existing benchmarks often limited by unimodal metrics that inadequately assess the intricacies of combined image-text outputs. To address these issues, we present RAG-IGBench, a thorough benchmark designed specifically to evaluate the task of Interleaved Generation based on Retrieval-Augmented Generation (RAG-IG) in open-domain question answering. RAG-IG integrates multimodal large language models (MLLMs) with retrieval mechanisms, enabling the models to access external image-text information for generating coherent multimodal content. Distinct from previous datasets, RAG-IGBench draws on the latest publicly available content from social platforms and introduces innovative evaluation metrics that measure the quality of text and images, as well as their consistency.  Through extensive experiments with state-of-the-art MLLMs (both open-source and proprietary) on RAG-IGBench, we provide an in-depth analysis examining the capabilities and limitations of these models. Additionally, we validate our evaluation metrics by demonstrating their high correlation with human assessments. Models fine-tuned on RAG-IGBench's training set exhibit improved performance across multiple benchmarks, confirming both the quality and practical utility of our dataset. Our benchmark is available at \textcolor{blue}{\url{https://github.com/USTC-StarTeam/RAG-IGBench}}.
\end{abstract}

\section{Introduction}
\begin{figure}
\centering
\includegraphics[width=14cm]{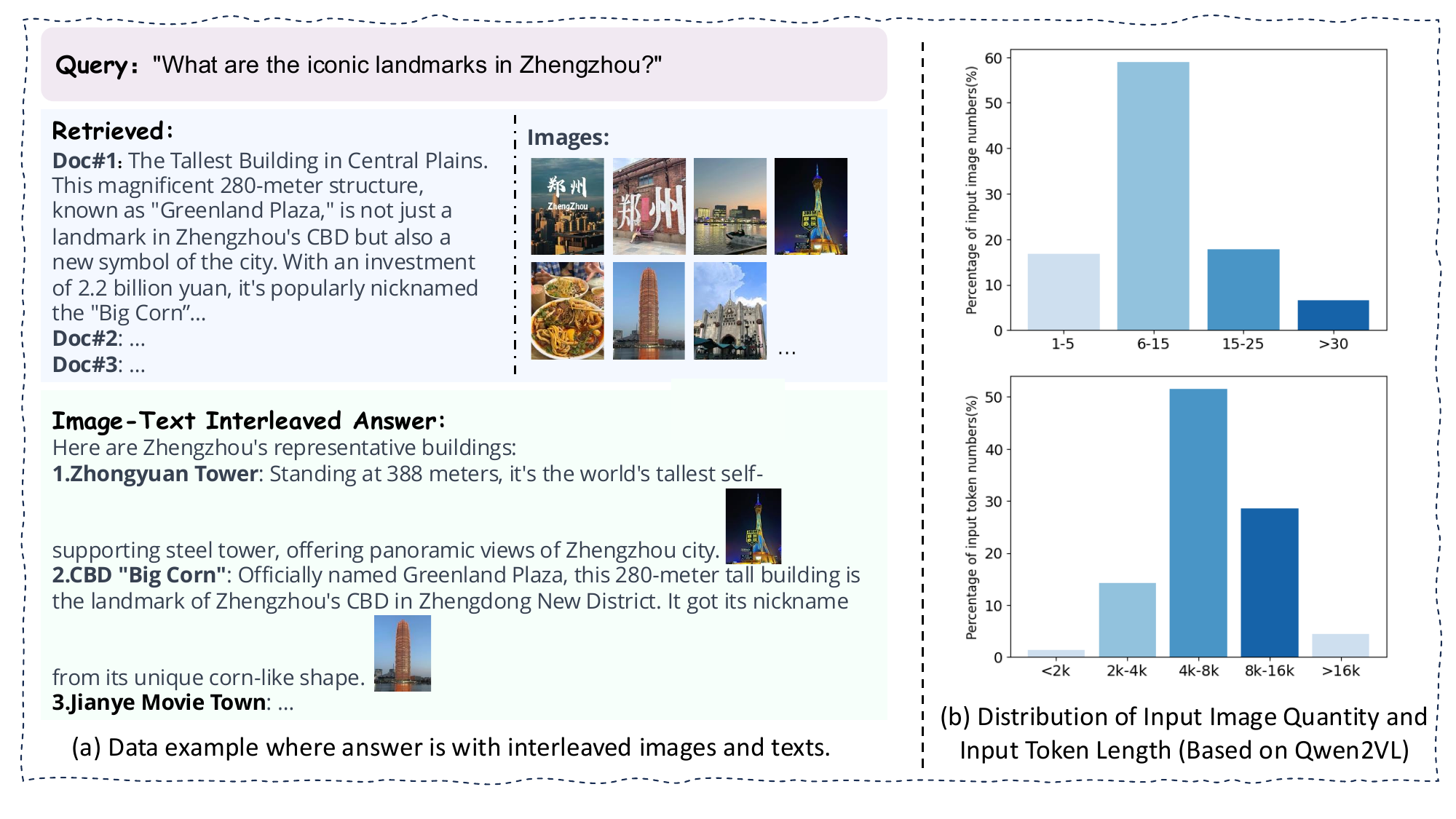}
\caption{Data example and statistics of input.}
\label{example}
\vspace{-0.8cm}
\end{figure}
Multimodal content generation has gained prominence in modern information systems, particularly for interleaved image-text generation in open-domain question answering. This emerging field addresses the need for responses combining textual explanations with visual elements for more comprehensive user interactions \cite{an2023openleaf, chen2024interleavedscenegraphinterleaved, tian2024mm}. The task aims to produce coherent sequences of interleaved text and illustrative images aligned with arbitrary queries \cite{koh2024generating, dong2023dreamllm, sun2024emugenerativepretrainingmultimodality, shen2024exploring}, supporting applications ranging from web content creation to visual storytelling \cite{huang2016visual}, recommendation \cite{wang2025generative, guo2024scaling, wu2024survey, liu2023user, zhang2025killing, ye2025fuxi}, and chain-of-thought explanations. Unlike previous approaches \cite{antol2015vqa, wang2016learning, hossain2019comprehensive, deng2021transvg, shen2025genkienhancingopendomainquestion, gu2025rapid, lv2025costeer} that generated single-modality responses, this integrated approach better addresses complex real-world needs.

Recent research \cite{an2023openleaf} has explored combining multimodal large language models (MLLMs) with diffusion models for coherent image-text generation. However, this approach suffers from semantic inconsistencies between modalities and reduced image quality due to independent generation processes. While novel unified transformer architectures \cite{chameleonteam2024chameleonmixedmodalearlyfusionfoundation, zhou2024transfusion, tian2024visual} attempt to handle image-text comprehension and generation simultaneously, they show limitations in following complex instructions, restricting their practical utility. Additionally, these transformer-based approaches require extensive datasets and computational resources, presenting further development challenges.

In addition, developing comprehensive benchmarks for interleaved image-text generation presents two significant challenges \cite{fu2024mme, shen2025optimizingsequentialrecommendationmodels, yin2024dataset, zhang2025td3}. First, despite the availability of extensive open-source image-text datasets \cite{changpinyo2021conceptual12mpushingwebscale, schuhmann2022laion, zhou2024lima}, their unlabeled nature and quality inconsistencies limit their applicability to benchmark evaluation, relegating them primarily to model pre-training. Second, current evaluation methodologies demonstrate considerable limitations \cite{liu2024holistic, xia2024mmie}. Conventional metrics such as Fréchet Inception Distance (FID) \cite{heusel2017gans} assess only image quality while neglecting image-text integration. Although recent MLLM-based evaluation approaches \cite{liu2024holistic, xia2024mmie} attempt to address this gap through specialized instructions, they introduce inherent model biases, inconsistencies from stochastic variations, and require task-specific manual adaptation of evaluation criteria.

To enhance image-text content generation for open-domain applications, we propose an innovative approach based on Retrieval-Augmented Generation (RAG) \cite{lewis2021retrievalaugmentedgenerationknowledgeintensivenlp}. As illustrated in Figure \ref{example}(a), our RAG-IG framework integrates MLLMs with retrieval mechanisms, enabling the processing of both user queries and retrieved multimodal content. The system generates markdown-formatted text with image placeholders, subsequently replaced with corresponding retrieved images to produce coherent interleaved image-text responses. This methodology significantly improves information clarity and presentation, yielding high-quality outputs that effectively integrate textual and visual elements while maintaining semantic consistency.

To address the evaluation challenges in image-text interleaved generation tasks, we propose a comprehensive benchmark framework that extends beyond current MLLM-based evaluation methods \cite{liu2024holistic, xia2024mmie}. Our evaluation framework systematically assesses generated content across three critical dimensions: textual quality, image quality, and image-text coherence. For textual assessment, we utilize ROUGE scores \cite{lin2004rouge} to evaluate the accuracy and fluency of generated text. For image evaluation, we propose modified metrics based on edit distance \cite{ristad1998learning} and Kendall correlation \cite{mcleod2005kendall} to assess both image selection accuracy and sequential arrangement. The image-text coherence is measured through two complementary metrics: CLIP-score \cite{hessel2021clipscore} and a novel semantic alignment score derived from embedding models. These metrics are then integrated to provide a holistic evaluation score for the generated content.

To establish a reliable benchmark, we have developed a systematic data collection and validation pipeline. The process begins with careful query selection and image-text content generation using our RAG-IG framework, followed by comprehensive human evaluation across multiple quality dimensions. We validate RAG-IG's effectiveness from both data and metric perspectives. First, models fine-tuned on our benchmark dataset demonstrate improved performance across various multimodal evaluation tasks, confirming the high quality of our data. Second, we validate our evaluation metrics through a systematic study of 200 sampled queries, where we compare our automated metrics against human assessments. The strong correlation between automated and human evaluations, measured by both Pearson \cite{sedgwick2012pearson} and Spearman \cite{sedgwick2014spearman} coefficients, demonstrates the robustness of our evaluation framework. These results establish our benchmark as a reliable evaluation tool for interleaved image-text content.

\begin{figure}
\centering
\includegraphics[width=14cm]{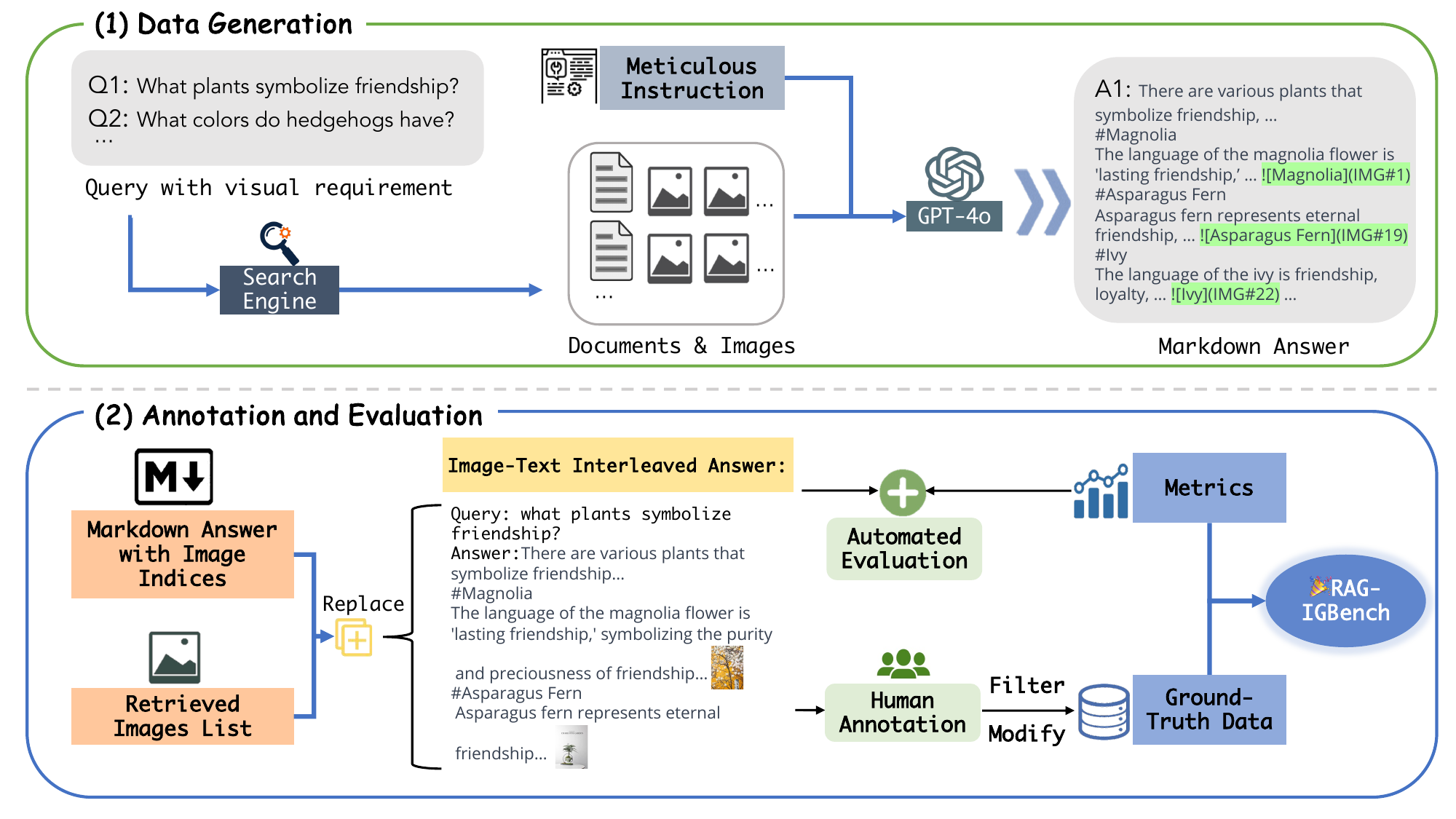}
\caption{Overview of the pipeline of RAG-IG and the data construction of RAG-IGBench}
\label{overview}
\vspace{-0.6cm}
\end{figure}

Our main contributions can be summarized as follows: (1) We introduce an innovative method for generating interleaved image-text content based on RAG. This approach provides multi-modal answers to queries requiring images, effectively integrating text and visuals. (2) We establish a novel benchmark for interleaved image-text generation with a systematically collected dataset and validate the high quality of our collection. (3) We propose innovative automated metrics for evaluation and demonstrate their reliability through correlation experiments between automated scores and human assessment results. (4) We perform extensive experiments on our benchmark, evaluating a range of mainstream state-of-the-art MLLMs, including both open-source and proprietary models. The results provide valuable insights for future research in this field.

\section{Related Work}
\textbf{Multimodal Large Language Models for Multimodal Generation.}
Recent advances in multimodal large language models (MLLMs) have substantially improved the integration of text and image modalities \cite{liu2024visual, zhu2023minigpt, dai2023instructblip, yin2024entropy}. These models have evolved through pipeline-based and end-to-end approaches. Pipeline-based approaches combine diffusion models \cite{rombach2022highresolutionimagesynthesislatent} for interleaved image-text generation \cite{koh2023generatingimagesmultimodallanguage} or utilize text-to-image retrieval to obtain visual information \cite{li2023blip, radford2021learning, yu2022coca}, while end-to-end approaches like Chameleon \cite{chameleonteam2024chameleonmixedmodalearlyfusionfoundation} and Show-o \cite{xie2024show} employ unified transformer architectures for joint image-text understanding and generation. Though these advances have enhanced models' capabilities in cross-modal understanding and generation, they face significant challenges in maintaining both high-quality image generation and image-text semantic alignment. RAG-based Interleaved Generation (RAG-IG) effectively addresses these limitations and requires fewer computational resources.

\textbf{Benchmarks for Multimodal Large Language Models.}
Current evaluation benchmarks for multimodal large language models (MLLMs) face several critical limitations. While large-scale datasets like MINT-1T \cite{koh2023generatingimagesmultimodallanguage} and Obelics \cite{laurenccon2024obelics} serve as valuable pre-training resources, their lack of annotations and limited image-text coherence restricts their effectiveness for evaluation. Traditional benchmarks \cite{hudson2019gqa}, though effective for basic capability assessment, fall short in evaluating complex reasoning abilities. Recent benchmarks \cite{li2023seed, zhang2024magicbrush, liu2024mmbench, yu2023mm, xia2024cares, wang2024mementos, tu2023many, cui2023holistic, huang2024chemeval} have addressed sophisticated reasoning evaluation, yet they primarily focus on isolated tasks, overlooking the critical aspect of interleaved comprehension and generation. Although advancing efforts like \cite{liu2024holistic}, \cite{xia2024mmie}, and \cite{zhou2025openingcomprehensivebenchmarkjudging} attempt to evaluate interleaved content through model-based methods, these benchmarks are either limited by their dataset scale or restricted by the limitations of their evaluation metrics. Our benchmark addresses these limitations by providing a comprehensive evaluation framework for interleaved multimodal comprehension and generation.

\textbf{Evaluation Metrics For Multimodal Generation.}
% Evaluation metrics for multimodal tasks face two primary challenges: assessing the quality of individual modalities and evaluating cross-modal coherence. Traditional single-modality metrics, such as BLEU \cite{papineni2002bleu} for text and FID \cite{heusel2017gans} for images, are limited to evaluating their respective modalities in isolation. While recent metrics like CLIPScore \cite{hessel2021clipscore} and X-IQE \cite{chen2023x} address text-image alignment, they focus solely on cross-modal correspondence without considering the intrinsic quality of each modality. Current model-based evaluation approaches, as demonstrated in INTERLEAVEDBENCH \cite{liu2024holistic} using GPT-based models and MMIE \cite{xia2024mmie} using fine-tuned smaller models, face inherent model biases and instability. Additionally, these evaluation frameworks require task-specific instruction adaptation, leading to potential evaluation inconsistencies across different tasks and significant manual effort.
Evaluating multimodal generation for open-domain question answering presents significant challenges \cite{an2023openleaf, Shao_2023_ICCV, wu2024openendedvisualqualitycomparison}, which we have systematically categorized into three key aspects. First, single-modality metrics (e.g., BLEU\cite{papineni2002bleu}, ROUGE\cite{lin2004rouge}  for text and FID \cite{heusel2017gans}, IS \cite{salimans2016improvedtechniquestraininggans} for images) fail to capture the interdependence between different modalities. Second, cross-modal alignment metrics like CLIPScore \cite{hessel2021clipscore} and X-IQE \cite{chen2023x} evaluate image-text correspondence but overlook the intrinsic quality of individual modalities. Third, recent model-based approaches face significant limitations: GPT-based scoring  \cite{liu2024holistic, zhang2023gpt4visiongeneralistevaluatorvisionlanguage} and fine-tuned models scoring \cite{xia2024mmie} introduce model-specific biases and suffer from evaluation instability \cite{bai2023benchmarkingfoundationmodelslanguagemodelasanexaminer, wang2024pandalmautomaticevaluationbenchmark}, and Interleaved Area approach \cite{zhou2025openingcomprehensivebenchmarkjudging} encounters considerable computational expense and limited scalability. Moreover, these frameworks require extensive task-specific adaptations, leading to inconsistent evaluation standards across different tasks and substantial manual effort in implementation.

\section{RAG-IGBench}
In this paper, we propose an approach for interleaved image-text generation based on Retrieval-Augmented Generation (RAG) and establish a comprehensive benchmark for systematic evaluation.  The following sections first present the formal task definition for RAG-based interleaved generation, followed by a detailed description of our RAG-IGBench, including data format, statistics, construction methodology, and evaluation metrics.

\begin{wraptable}{r}{7cm}
\centering
\captionof{table}{Dataset statistics. The token statistics are calculated based on the tokenizer from Qwen2VL.}
\resizebox{0.5\textwidth}{!}{
\begin{tabular}{lrr}
     \noalign{\hrule height 1pt}
     \textbf{Statistic} & \textbf{Number} & \textbf{Percentage}\\
     \hline
     Total queries & 6,057 & - \\
     %  \hspace{1em}- Health & 464 & 14.69\% \\
     %  \hspace{1em}- Finance & 95 & 3.01\% \\
     %  \hspace{1em}- Education & 258 & 8.17\% \\
     %  \hspace{1em}- Entertainment & 349 & 11.05\% \\
     %  \hspace{1em}- Fashion & 340 & 10.77 \% \\
     %  \hspace{1em}- Family & 360 & 11.40\% \\
     %  \hspace{1em}- Technology & 491 & 11.55\% \\
     %  \hspace{1em}- Culture & 801 & 25.36\% \\
     % \hline
     Questions with images & 6,057 & 100\% \\
     Questions with answer label & 6,057 & 100\% \\
     Avg./Max. Retrieved images & 12.96/51 & - \\
     Avg./Max. Generated images & 2.26/10 & - \\
     Avg./Max. Input tokens & 7,914.22/39,959 & - \\
     Avg./Max. Answer tokens & 291.40/2,967 & - \\
     \noalign{\hrule height 1pt}
\end{tabular}
}
\label{data stas}
\vspace{-1.0em}
\end{wraptable}

\vspace{-0.3cm}
\subsection{Task Definition}
\vspace{-0.2cm}
The task of interleaved image-text generation aims to produce responses that effectively combine textual and visual information for a given query.
We adopt the RAG framework \cite{lewis2021retrievalaugmentedgenerationknowledgeintensivenlp} to enhance the coherence and quality of the generated content.  As illustrated in Figure \ref{overview}, given a query \textit{q}, we begin by retrieving a set of relevant documents \( D = \{ d_{1},d_{2},\ldots,d_{n} \} \) and their corresponding images \( I = \{ I_{1},I_{2},\ldots,I_{n} \} \) where \textit{n} is the number of retrieved documents and \( I_{i} \) is the image set of the \textit{i-th} document. Then, the retrieved content, in conjunction with the query and our elaborate instruction, is utilized as input for multi-modal large language models (MLLMs). Each image in the input is distinctly labeled with an index number. The MLLM generates a response in markdown format, where images are represented as ![image description](IMG\#k), with k indicating the image index. We then replace these indices with corresponding image URLs to produce the final multimodal response.

However, implementing the RAG-IG paradigm with multiple images poses significant challenges: large image sets expand the input context window substantially, straining MLLM performance and computational resources \cite{song2024milebench, huang2024opera}. Therefore, we retrieve a maximum of three documents per query. Additionally, we exclude the queries that can be adequately addressed through text-only responses.

\vspace{-0.3cm}
\subsection{Dataset Format and Statistics}
\vspace{-0.2cm}
Each sample in our dataset is structured as a tuple (\textit{q}, \textit{gt}, \textit{D}, \textit{I}), where \textit{q} represents the original user query, \textit{gt} denotes the ground-truth answer in markdown format incorporating both textual and visual elements, \textit{D} represents the retrieved contextual documents and \textit{I} represents their corresponding images. As shown in Table \ref{data stas}, \benchmark consists of 6,057 curated samples where each query is selected based on explicit visual requirements.

% \begin{minipage}{0.54\linewidth}
% \centering
%     \subfloat[Input Image Numbers]{\includesvg[width=.45\linewidth] {sections/images/input_img_num.svg}}\hspace{5pt}
%     \subfloat[Answer Image Numbers]{\includesvg[width=.45\linewidth]{sections/images/generate_img_num.svg}}\\
%     % \subfloat[Input Tokens]{\includesvg[width=.45\linewidth]{sections/images/input_token_num.svg}}\hspace{5pt}
%     % \subfloat[Answer Tokens]{\includesvg[width=.45\linewidth]{sections/images/answer_token_num.svg}}
%     \caption{Distribution of Input Image Numbers, Input Tokens, Answer Image Numbers and Answer Tokens. The token statistics are calculated based on the tokenizer from Qwen2VL.}
%     \label{image num}
% \end{minipage}

\vspace{-0.3cm}
\subsection{Dataset Construction}\label{3.3}
\vspace{-0.2cm}
The process of constructing benchmark data comprises three key stages, each of which plays a crucial role in ensuring the high quality of the final dataset. 

In the first stage, we guide MLLMs (including GPT4o\cite{hurst2024gpt}, Claude-3.5-Sonnet\cite{TheC3}, et al.) with meticulously crafted instructions to generate the raw question-answer data. This stage takes a query, retrieved documents, and relevant images as input into an MLLM, resulting in a Markdown-formatted answer that includes image indices. 
%All query-relevant documents and images were sourced from \textit{Xiaohongshu}, a popular Chinese social media platform.
To improve answer correctness and ensure consistency between text and images, we took inspiration from the Chain of Thought (CoT) \cite{wei2023chainofthoughtpromptingelicitsreasoning} and the ICL approach \cite{dong2024surveyincontextlearning}, which begins by instructing the model to analyze the input query, categorizing it into one of four types: "what-is", "how-to", "yes-or-no", and "head-to-head". The main reason for categorizing queries is that different types of queries have varying requirements for visual information. Please refer to the appendix \ref{detailed_anno} and \ref{prompt} for detailed information and the specific generation prompt.

In the second stage, we annotate first-stage results through a rigorous process that is both time-consuming and labor-intensive. We first filter out cases lacking image indices in the answers. Annotators then evaluate the remaining data based on text quality, image quality, image-text consistency, and overall quality. To establish high-quality ground-truth answers, we conduct thorough image selection refinement, manually adding, removing, or replacing images in answers according to the candidate images. All annotators are multimodal assessment experts with high inter-rater reliability scores [\( \kappa \) > 0.85], ensuring precise ground-truth answers.

In the third stage, we filter the QA pairs based on the annotation results of the second stage. QA pairs with low scores in any dimension are excluded from the dataset. Besides, we retain cases that have been successfully improved through manual refinement. This systematic filtering process results in our final benchmark dataset.

Through this three-stage process, we constructed our final \benchmark. Answers are stored in markdown format with image indices rather than direct URLs, allowing for systematic evaluation and easy conversion to complete multimodal responses by mapping indices to their corresponding image URLs. It should be noted that the data was systematically reviewed for privacy concerns and any content containing potentially identifiable personal information was excluded.

\begin{table}
    \centering
    \caption{Comparisons between Benchmark and existing open-sourced multi-modal evaluation benchmarks.\vspace{0.2cm}}
    \resizebox{\textwidth}{!}{
    \begin{tabular}{lccccc}
        \toprule
         \textbf{Dataset Name} & \textbf{\makecell[c]{Multi-Images \\ Input}} & \textbf{\makecell[c]{Interleaved \\ Input}} & \textbf{\makecell[c]{Interleaved \\ Generation}} & \textbf{\makecell[c]{Average \\ Images}} & \textbf{Metric}\\ 
         \midrule
         MIRB \cite{zhao2024benchmarking} & \textcolor{ForestGreen}{\ding{51}} & \textcolor{red}{\ding{55}} & \textcolor{red}{\ding{55}} & 3.78 & ACC    \\
         MEGA-Bench \cite{chen2024mega} & \textcolor{ForestGreen}{\ding{51}} & \textcolor{red}{\ding{55}} & \textcolor{red}{\ding{55}} & 2.0 & 45 metrics    \\
         Qbench2 \cite{wu2023q} & \textcolor{ForestGreen}{\ding{51}} & \textcolor{red}{\ding{55}} & \textcolor{red}{\ding{55}} & 2.0 &  ACC \\
         NLVR2 \cite{suhr2019corpusreasoningnaturallanguage} & \textcolor{ForestGreen}{\ding{51}} & \textcolor{red}{\ding{55}} & \textcolor{red}{\ding{55}} & 2.0 &  ACC \\
         BLINK \cite{fu2024blinkmultimodallargelanguage} & \textcolor{ForestGreen}{\ding{51}} & \textcolor{red}{\ding{55}} & \textcolor{red}{\ding{55}} & 1.93 &  ACC \\
         MuirBench \cite{wang2024muirbench} & \textcolor{ForestGreen}{\ding{51}} & \textcolor{ForestGreen}{\ding{51}} & \textcolor{red}{\ding{55}} & 4.3 &  ACC  \\
         SEED-Bench-2 \cite{li2024seed} & \textcolor{ForestGreen}{\ding{51}} & \textcolor{ForestGreen}{\ding{51}} & \textcolor{red}{\ding{55}} & 2.74 & ACC    \\
         VL-ICL Bench \cite{zong2024vl} & \textcolor{ForestGreen}{\ding{51}} & \textcolor{ForestGreen}{\ding{51}} & \textcolor{red}{\ding{55}} & 1.26 & Acc \& MLLM    \\
         MMMU \cite{yue2024mmmu} & \textcolor{ForestGreen}{\ding{51}} & \textcolor{ForestGreen}{\ding{51}} & \textcolor{red}{\ding{55}} & 1.16 &  ACC \\
         INTERLEAVEDBENCH\cite{liu2024holistic} & \textcolor{ForestGreen}{\ding{51}} & \textcolor{ForestGreen}{\ding{51}} & \textcolor{ForestGreen}{\ding{51}} & 1.7 & GPT-4o based  \\
         MMIE \cite{xia2024mmie} & \textcolor{ForestGreen}{\ding{51}} & \textcolor{ForestGreen}{\ding{51}} & \textcolor{ForestGreen}{\ding{51}} & 8.5 & Fine-tuned VLM based \\
        \midrule
         RAG-IGBench (Ours) & \textcolor{ForestGreen}{\ding{51}} & \textcolor{ForestGreen}{\ding{51}} & \textcolor{ForestGreen}{\ding{51}} & \textbf{13.0} & Multi-dimensional metrics  \\
        \bottomrule
    \end{tabular}}
    \label{comparision}
\vspace{-0.5cm}
\end{table}
\vspace{-0.3cm}
\subsection{Evaluation Metric}\label{3.4}
\vspace{-0.2cm}
This section presents our evaluation metrics for multimodal responses. Following the annotation criteria established in \ref{3.3}, we evaluate responses across three dimensions: text quality, image quality, and image-text alignment. Unlike previous works that rely on MLLM-based scoring methods such as INTERLEAVEDBENCH \cite{liu2024holistic} and MMIE \cite{xia2024mmie}, we implement rule-based metrics leveraging ground-truth references, which ensures unbiased and reproducible evaluation results. The following subsections detail our specific metrics for each dimension.
% This section presents the evaluation metrics of our benchmark. Following the same criteria used in the manual annotation process in \ref{3.3}, we evaluate model-generated multimodal responses across three dimensions: text quality, image quality, and image-text alignment. With ground-truth answers now available as references, we implement rule-based evaluation metrics rather than the MLLM scoring methods used in INTERLEAVEDBENCH\cite{liu2024holistic} and MMIE\cite{xia2024mmie}. This approach eliminates potential model biases inherent in MLLM-based evaluation and ensures consistent, reliable results. The following sections describe our specific metrics for assessing text quality, image quality, and image-text alignment.

\textbf{Text Quality}: In the field of natural language processing (NLP), there are a variety of effective metrics to evaluate the quality of the generated text, including BLEU \cite{papineni2002bleu}, ROUGE \cite{lin2004rouge}, and perplexity. In \benchmark, we focus on measuring the semantic alignment between the generated text and the ground truth answers. We adopt the ROUGE-1 score as our evaluation metric due to its effectiveness in capturing lexical overlap and semantic similarity between text passages.  Although we also evaluated the ROUGE-2 and ROUGE-L metrics, empirical testing showed that ROUGE-1 provides better discriminative power to distinguish the quality of generated responses.

\textbf{Image Quality}: 
Previous image quality assessment studies have relied on metrics like the Fréchet Inception Distance (FID) \cite{heusel2017gans} and the Inception Score (IS) \cite{barratt2018note}, which measure statistical similarity between the generated and real distributions of characteristic features of the image. However, our RAG-IG framework selects images rather than generates them, making these traditional metrics inapplicable. Since both generated and ground-truth answers contain images selected from the same candidate pool, our evaluation focuses on two aspects: the selection accuracy and the ordering correctness of images. We formulate this as an ordered list comparison problem, employing Edit Distance to measure selection accuracy and Kendall Score to evaluate ordering consistency.

Specifically, let \( I_{generated} = \{ img_{1}, img_{2}, \ldots, img_{m} \} \) and \( I_{gt} = \{ img_{1}, img_{2}, \ldots, img_{n} \} \) where \( I_{generated} \) and \( I_{gt} \) are lists composed of images from the generated answer and ground truth, respectively, and \textit{m} and \textit{n} represent the lengths of two lists.

For the Edit Distance, we measure the similarity between the generated image sequence \( I_{generated} \) and \( I_{gt} \) by calculating the minimum number of operations (insertion, deletion, and substitution) required to transform one into the other. This computation can be efficiently implemented using dynamic programming. To handle varying sequence lengths, we normalize the score as:
\begin{equation}
    \setlength\abovedisplayskip{3pt}
    \setlength\belowdisplayskip{3pt}
    \text{Edit Distance = }1-\frac{dp(m, n)}{max(m,n)}
\end{equation}

where \( dp \) is the dynamic programming matrix. This normalization ensures comparable scores across sequences of different lengths, with higher values indicating greater similarity.

For the Kendall Score, inspired by the Kendall rank correlation coefficient \cite{mcleod2005kendall}, commonly known as Kendall's \( \tau \) coefficient, we calculate it by counting the proportion of concordant pairs between the two lists relative to all correct pairs in \( I_{generated} \). To elaborate, let \( I_{correct}=\{img_{1},img_{2},\ldots,img_{o}\} \) be the intersection of two image lists, \( I_{generated} \) and \( I_{gt} \), where 
\textit{o} denotes the length of the intersection, and the images in \( I_{correct} \) follow the order in \( I_{generated} \). Then define \( index(img_{k}, I_{gt}) \) refer to the index of the image \( img_{k} \) within \( I_{gt} \). For each possible pair \( (img_{i},img_{j}) \) in \( I_{correct} \), where \( i<j \leq o \), determine if it is a concordant pair by checking if \( index(img_{i}, I_{gt})<index(img_{j}, I_{gt}) \). At last, we count the ratio of the number of concordant pairs to the total number of possible pairs formed by elements in \( I_{correct} \) as the final Kendall Score. Therefore, we can derive the following formula for the Kendall Score:

\begin{small}
\begin{equation}
    \text{Kendall Score}\!=\left\{
    \begin{aligned}
    &\!\frac{\sum\limits_{1\leq i<j\leq o}\mathbb{I}(index(img_i,I_{gt})\!<\!index(img_j,I_{gt}))}{\frac{1}{2}o(o\!-\!1)},   & o>1 \\
    &\frac{o}{max(m,n)},  &o\leq 1 
    \end{aligned}
    \right.
\end{equation}
\end{small}

This dual-metric enables a comprehensive evaluation that considers both the selection of appropriate images and their proper arrangement in the response.

\textbf{Image-Text Consistency}: In our \benchmark, we first employ CLIP Score \cite{hessel2021clipscore} to assess image-text consistency by measuring cosine similarity in a shared vector space. However, it has notable limitations: it struggles with complex semantics, inherits biases from the pre-training data, and captures limited context for individual images. To overcome these limitations, we propose the Alignment Score metric, which evaluates textual similarity between the context surrounding the correct image in the generated answer and its corresponding context in the ground-truth answer. This approach better captures the contextual relationship in interleaved multimodal content. It can be formulated as follows:

\begin{equation}
    \text{Alignment Score}=\frac{\sum_{i=1}^{o} \operatorname{Sim}\left(C_{img_{i}}, G_{img_{i}}\right)}{o}
\end{equation}

Here \( img_{i}\in I_{correct} \), where \( C_{img_{i}} \) denotes the context of \( img_{i} \) in generated answer and \( G_{img_{i}} \) represents the context of \( img_{i} \) in ground-truth answer. This approach is predicated on the notion that the same image should be embedded in similar contextual settings. We evaluate image-text alignment by comparing both the selected images and their contextual usage with the ground-truth answer. Strong alignment between the generated and ground-truth answers typically indicates proper integration of images within their textual context.

The CLIP Score and Alignment Score provide complementary measures for evaluating image-text consistency, with the former assessing direct semantic alignment and the latter measuring contextual similarity between image usage patterns.

\vspace{-0.3cm}
\subsection{Comparison With Other Multi-modal Benchmarks}
\vspace{-0.2cm}
As illustrated in the table \ref{comparision}, \benchmark distinguishes itself from existing multimodal benchmarks in several significant ways: \textbf{(1) Multi-modal Multi-Image Input:} \benchmark requires MLLMs to process multiple images alongside text simultaneously, presenting a more complex challenge than existing benchmarks that typically handle single-modality input or limited image counts; \textbf{(2) Open-Domain Interleaved Generation:} Unlike INTERLEAVEDBENCH \cite{liu2024holistic} and MMIE \cite{xia2024mmie}, which focus on specific tasks like step-by-step instructions, our benchmark is designed for open-domain queries requiring interleaved image-text responses, enabling comprehensive and visually enriched answers across various domains; \textbf{(3) Novel Comprehensive Evaluation Metrics:} We introduce multi-dimensional evaluation metrics specifically designed for interleaved generation tasks, addressing three key dimensions: text quality, image quality, and image-text consistency (detailed in \ref{3.4}). Our \benchmark overcomes the limitations of existing metrics that either lack task specificity or rely on MLLM-based evaluation, which can introduce potential bias and instability.

\begin{table}[htb]
\vspace{-1.0em}
    \centering
    % \captionsetup{width=0.9\linewidth}
    \caption{Evaluation results of mainstream MLLMs on RAG-IGBench. Bold and underlined entries represent the best performance in proprietary and open-source models, respectively.\vspace{0.2cm}}
    
    \resizebox{\textwidth}{!}{
    \begin{tabular}{lccccc>{\columncolor{gray!20}}c}
        \toprule
        \textbf{Models} & \textbf{Text Quality} & \multicolumn{2}{c}{\textbf{Image Quality}}  & \multicolumn{2}{c}{\textbf{Image-Text Consistency}}  & \textbf{Mean}\( \uparrow \) \\
        \midrule
         & \textbf{Rouge-1}\( \uparrow \) & \textbf{\makecell[c]{Edit \\ Distance}}\( \uparrow \) & \textbf{\makecell[c]{Kendall \\ Score}}\( \uparrow \) & \textbf{\makecell[c]{Alignment \\ Score}}\( \uparrow \) & \textbf{\makecell[c]{CLIP \\ Score}}\( \uparrow \) & \\
        \midrule
        \multicolumn{7}{c}{Proprietary MLLMs}\\
        \midrule
        GPT4o & \textbf{57.42} & \textbf{51.28} & \textbf{46.50} & \textbf{38.81} & \textbf{36.04} & \textbf{46.01}  \\
        Claude3.5-sonnet & 35.98 & 29.98 & 21.91 & 30.68 & 35.56 & 30.81  \\
        Gemini-1.5-pro & 46.35 & 42.22 & 34.57 & 35.07 & 34.85 & 38.61 \\
        QwenVL-Max & 49.24 & 44.66 & 38.02 & 34.55 & 38.28 & 40.95 \\
        \midrule
        \multicolumn{7}{c}{Open-source MLLMs}\\
        \midrule
        Qwen2VL-7B & 43.21 & 22.23 & 18.93 & 18.77 & 27.84 & 26.20 \\
        Qwen2VL-72B & 49.49 & 36.66 & \underline{31.40} & 26.49 & 32.69 & 35.34 \\
        Llava Onevision 72B  & 42.89 & 24.77 & 19.66 & 18.20 & 27.19 & 26.54  \\
        InternVL2.5 8B & 41.98 & 24.53 & 19.20 & 21.94 & 28.53 & 27.24  \\
        InternVL2.5 78B & \underline{50.71} & 36.86 & 27.10 & \underline{35.68} & 33.23 & 36.71 \\
        NVLM-D-72B & 38.43 & 13.97 & 11.87 & 8.57 & 13.82 & 17.33 \\
        InternVL2-Llama3-76B & 43.91 & 25.18 & 18.15 & 25.38 & 27.66 & 28.08 \\
        Qwen2.5VL-7B & 44.79 & 24.56 & 19.75 & 19.87 & 25.45 & 26.88 \\ 
        Qwen2.5VL-72B & 43.18 & \underline{40.38} & 30.12 & 35.26 & \underline{40.03} & \underline{37.79} \\ 
        \bottomrule
    \end{tabular}}
    \label{main result}
\vspace{-1.0em}
\end{table}

\section{Experiment}\label{exp}

This section describes our experimental setup and presents the evaluation results of mainstream MLLMs on our \benchmark, followed by a detailed analysis of model behaviors.

\subsection{Experiment Setup}

\textbf{Evaluation Models.} In this study, we evaluated the mainstream state-of-the-art MLLMs on the \benchmark including four closed-source models (GPT-4o \cite{hurst2024gpt}, Claude 3.5-Sonet \cite{TheC3}, Gemini-1.5 \cite{team2023gemini}, QwenVL-Max \cite{bai2023qwen}) and six open-source models (Qwen2vl-7B/72B \cite{wang2024qwen2vlenhancingvisionlanguagemodels}, Llava OneVision 72B \cite{li2024llavaonevisioneasyvisualtask}, InternVL2.5 8B/78B \cite{chen2024expanding}, NVLM-D-72B \cite{nvlm2024}, InternVL2-Llama3-76B \cite{chen2024internvl}, Qwen2.5VL 7B/72B \cite{qwen2.5-VL}). The RAG-IG framework requires models to process multimodal inputs and multiple images simultaneously, which excludes certain models like LlaVA-1.6 \cite{liu2024visual} despite their strong image-text understanding capabilities. Furthermore, the evaluated models must strictly follow instructions to generate valid markdown results that can be transformed into image-text interleaved answers. Due to these requirements, models with limited parameters and relatively weaker image-text understanding capabilities were not included in our evaluation.  All models employed greedy decoding for answer generation, and all evaluations were conducted on a single machine equipped with 8 H800 GPUs.

\textbf{Implementation Details.} We utilize CLIP-ViT-Large-Patch14 \cite{hessel2021clipscore} from OpenAI for CLIP Score calculation and Conan-embed-ding \cite{li2024conanembeddinggeneraltextembedding} for Alignment Score calculation. The choice of Conan-embedding is based on its superior performance in text embedding tasks. Besides, all images were constrained to a uniform width of 540 pixels for the main experiments. To facilitate a more systematic comparison across different models, we normalized both metrics to a standardized scale.

\begin{table}
    \centering
    \caption{Correlation results with Human Evaluation of Image Quality, Image-Text Consistency and Overall Assessment.\vspace{0.2cm}}  
    \resizebox{.8\textwidth}{!}{
    \begin{tabular}{l|cccccc}
        \toprule
        \multirow{2}*{\textbf{Metric}} & \multicolumn{2}{c}{\textbf{Image Quality}} & \multicolumn{2}{c}{\textbf{Image-Text Consistency}} & \multicolumn{2}{c}{\textbf{Overall Assessment}} \\
        \cmidrule(r){2-3} \cmidrule(r){4-5} \cmidrule(r){6-7}
        ~ & \textbf{Pearson} & \textbf{Spearman} & \textbf{Pearson} & \textbf{Spearman} & \textbf{Pearson} & \textbf{Spearman} \\
        \midrule
        FID & -0.223 & -0.217 & - & - & - & - \\
        IS & -0.174 & -0.139 & - & - & - & - \\ 
        CLIP-Score  & - & - & -0.108 & -0.094 & - & - \\ 
        GPT4o based & 0.097 & 0.112 & 0.151 & 0.104 & 0.017 & 0.028 \\
        RAG-IGBench & \textbf{0.256} & \textbf{0.244} & \textbf{0.172} & \textbf{0.145} & \textbf{0.143} & \textbf{0.152} \\
        \bottomrule
    \end{tabular}
    }
    \label{Correlation results}
\vspace{-1.0em}
\end{table}

\subsection{Main Results}

We present our primary evaluation results on RAG-IG in Table \ref{main result}, with proprietary models in the upper portion and open-source models in the lower portion. Our analysis reveals several key findings: (1) In terms of overall performance, GPT4o leads among all models, followed by Gemini-1.5-pro. Among open-source models, Qwen2.5VL-72B achieves the highest performance, closely followed by InternVL2.5 78B. Claude-3.5-sonnet demonstrates relatively lower performance, primarily due to incorrect images in its responses. (2) The performance gap between proprietary and open-source models varies significantly across different metrics. The small gap in CLIP Score between proprietary and open-source models suggests that current MLLMs have reached comparable proficiency in image-to-text tasks. However, larger performance disparities appear in the image quality dimension, which requires image-to-text matching capabilities. We attribute these significant differences to the limited exploration of such image-to-text matching tasks in previous multimodal research.

Analysis of Table \ref{main result} reveals a substantial performance gap between Qwen2VL-7B and Qwen2VL-72B. This disparity is not only in text quality but also in image quality and image-text alignment, demonstrating the significance of scaling up MLLM architectures.

\begin{wraptable}{r}{8.5cm}
\vspace{-0.3cm}
    \captionof{table}{Performance of Qwen2VL models benefits from fine-tuning on RAG-IGBench training set, which demonstrates the high quality of our dataset.}
    \label{sft result}
    \centering
    \resizebox{0.6\textwidth}{!}{
    \begin{tabular}{lccc}
        \toprule
        \textbf{Benchmark} & Qwen2VL-7B & Qwen2VL-7B-sft & Variation \\
        \midrule
        RAG-IGBench(Dev) & 26.12 & 36.10 & \textcolor{ForestGreen}{+9.98} \\
        MMB \cite{liu2024mmbench} & 81.62 & 82.04 & \textcolor{ForestGreen}{+0.42} \\
        BLINK \cite{fu2024blinkmultimodallargelanguage} & 53.34 & 54.50 & \textcolor{ForestGreen}{+1.16}\\
        Mantis-Eval \cite{jiang2024mantisinterleavedmultiimageinstruction} & 49.06 & 57.55 & \textcolor{ForestGreen}{+8.49}\\
        Q-Bench2 \cite{wu2023q} & 61.80 & 68.60 & \textcolor{ForestGreen}{+6.80}\\
        NLVR2 \cite{suhr2019corpusreasoningnaturallanguage} & 84.01 & 80.56 & \textcolor{red}{-3.45}\\
        \midrule
        Avg. & 59.33 & 63.23 & \textcolor{ForestGreen}{+3.90} \\
        \bottomrule
    \end{tabular}}

\end{wraptable}

Additionally, among open-source models, those beyond the Qwen-2VL series and InternVL2.5 demonstrate notably inferior performance. We attribute the superior performance of Qwen2VL and InternVL2.5 to their specialized image processing approaches: Qwen2-VL employs mROPE for adaptive resolution handling, while InternVL2.5 implements a dynamic resolution strategy, processing input images into tiles of \( 448 \times 448 \) pixels. This represents a promising direction for future MLLM advancement.

% \vspace{-0.3cm}
\subsection{Analysis}\label{4.3}
% \vspace{-0.2cm}
In this subsection, we conduct a detailed analysis of the results, centering on four key research questions related to RAG-IGBench:

\begin{figure}[H]
\begin{minipage}{0.45\linewidth}
        \centering
		\includegraphics[width=0.99\columnwidth]{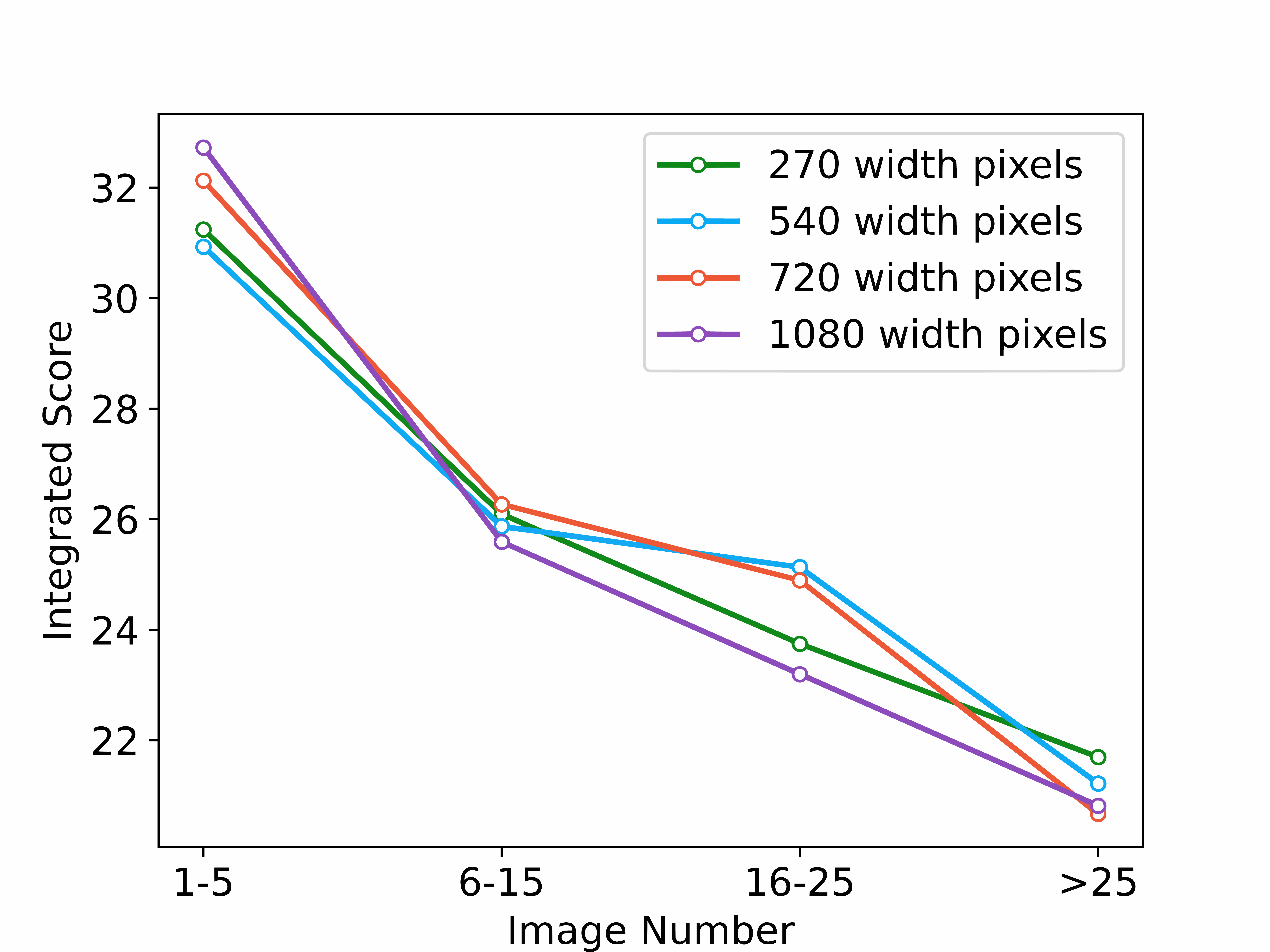}
		\caption{Ablation experiment result of different resolutions and amounts of input images.}
		\label{ablation}
\end{minipage}
\begin{minipage}{0.54\linewidth}
\centering
\captionof{table}{Statistics of invalid format and hallucinations generated by MLLMs}
\resizebox{\textwidth}{!}{
\begin{tabular}{lcc}
\toprule
    \textbf{Models} & \makecell[c]{Invalid \\ format} & Hallucination \\
    \midrule
    Qwen2VL-7B & 5 & 6 \\
    Qwen2VL-72B & 0 & 5 \\
    Llava Onevision 72B & 1579 & 12  \\
    InternVL2.5 78B & 0 & 14 \\
    NVLM-D-72B & 45 & 39 \\
    InternVL2-Llama3-76B & 44 & 94 \\
    \bottomrule
\end{tabular}
}
\label{hallu and invalid}

\end{minipage}
\end{figure}

\textbf{RQ1.} How is the consistency of our metrics with human evaluation?

\textbf{RQ2.} How is the data quality of RAG-IGBench?

\textbf{RQ3.} What challenges encountered by MLLMs in RAG-IG?

\textbf{RQ4.} How does the amount and resolution of input images influence the performance of MLLMs?

% \vspace{-0.3cm}
% \subsubsection{\textbf{The consistency of our metrics with human evaluation}}
% \vspace{-0.2cm}
\textbf{The consistency of our metrics with human evaluation. }
We evaluated our RAG-IGBench metrics through correlation studies with 200 randomly sampled cases. Our metrics assess image quality (Edit Distance and Kendall Score), image-text consistency (Alignment Score and CLIP Score), and overall assessment. We computed Pearson and Spearman correlations between these metrics and human evaluations, comparing them with conventional metrics (FID, IS, et al.) and GPT4o-based evaluation. Table \ref{Correlation results} shows that our metrics outperformed conventional metrics and GPT4o-based evaluation in terms of correlation with human evaluation.

% \vspace{-0.3cm}
% \subsubsection{\textbf{The data quality of RAG-IGBench}}
% \vspace{-0.2cm}
\textbf{The data quality of RAG-IGBench. }
To illustrate the high quality of our dataset, we conducted a systematic evaluation by partitioning RAG-IGBench into training and test sets and subsequently fine-tuning the Qwen2VL series model on the training set. The experimental results, as shown in the table \ref{sft result}, demonstrate not only significant performance improvements on the test set but also notable gains across various existing multi-modal benchmarks. Particularly noteworthy is the enhanced performance on benchmarks that require multi-image comprehension capabilities, such as BLINK \cite{fu2024blinkmultimodallargelanguage}. These comprehensive improvements across different evaluation metrics underscore the quality of our dataset.

% \vspace{-0.3cm}
% \subsubsection{\textbf{Challenges encountered by MLLMs in RAG-IG}}
% \vspace{-0.2cm}
\textbf{Challenges encountered by MLLMs in RAG-IG. }
RAG-IG poses substantial challenges to MLLMs, particularly in their ability to comprehend and process interleaved image-text inputs containing multiple images. Through systematic error analysis, we find that not all queries produce valid outputs, and we have categorized these failure cases into three distinct categories:
(1) Hallucination: Our experiments reveal that models frequently hallucinate image index numbers, generating indices that exceed the number of retrieved images (see Figure \ref{error case}). 
(2) Instruction following ability: Despite explicit formatting instructions in prompt, models often generate syntactically incorrect Markdown structures.
(3) Length of context windows: The combination of multiple images and retrieved texts results in extensive context lengths, thereby impacting model performance and memory efficiency. Consequently, we have to exclude samples with excessive images and restrict the maximum number of retrieved documents to three.
Table \ref{hallu and invalid} presents statistics on hallucination and format errors across model variants. We focus on open-source MLLMs where these issues are prevalent, as proprietary models show minimal errors. 

% \vspace{-0.3cm}
% \subsubsection{\textbf{Ablation study on amount and resolution of input images}}
% \vspace{-0.2cm}
\textbf{Influence of the amount and resolution of input images. }
We conducted systematic ablation studies on Qwen2VL-7B, varying the number and resolution of input images (Figure \ref{ablation}). Our quantitative observations indicate the following:
(1) Model performance decreases with more input images, even at low resolutions, indicating MLLMs' inherent limitations in processing multiple images beyond token-length constraints.
(2) Higher image resolutions improve performance with 1-5 images but degrade performance with more images, suggesting a critical trade-off between detail richness and context length management.

% We conduct ablation experiments using QwenVL2-7B with varying numbers and resolutions of input images. The detailed results are presented in Figure \ref{ablation}. Our observations indicate the following: 
% (1) As the number of input images increases, the model's performance consistently declines, underscoring MLLMs' current limitations in handling multiple-image inputs. This decline occurs even at low image resolutions, suggesting that the issue is not merely due to an increase in tokens from additional images but also related to MLLMs' inherent limitations in processing multiple images. 
% (2) When the input includes only 1-5 images, increasing image resolution significantly enhances model performance. However, with more images, the model performs worse at the higher resolution. We attribute this to the substantial increase in context length required to process higher-resolution images. While higher resolutions can provide finer image details, the corresponding longer context sequences pose significant challenges to the model's capacity, ultimately affecting its overall performance.

% We believe these empirical results can provide valuable insights into the scalability and efficiency considerations in multi-modal systems, particularly regarding the balance between input complexity and model performance.

\subsection{Other Text Quality Metrics}\label{4.4}
% 我们选择rouge-1作为文本评估指标的原因是它作为广泛应用的文本指标的同时，计算的复杂度较低，能够实现高效的评估，但考虑到近年来不断有更stronger的文本指标被提出，如CLAIR，SPICE等，因此我们也在表中报告了几个主流模型在这些指标下的评估结果用于参考。其中，CLAIR指标利用GPT-4o作为评估模型，
% The CIDEr and SPICE metrics generally require multiple reference sentences (i.e., multiple ground truths) to produce reliable results. Since RAG-IGBench provides only a single reference sentence, the reliability of these metrics is compromised.  CLAIR demonstrates good capability in assessing text quality differences among various models. Nevertheless, we need to account for GPT-4o's inherent model bias, potential evaluation instability caused by API changes, and the associated additional expenses.与CLAIR相比，rouge-1指标不需要额外的llm，基于规则评估文本质量，能进行更加高效的评估。
% 综上所诉，ROUGE指标在评估可靠性上具有局限性，而CLAIR指标在引入模型偏见和额外的开销上有局限性，综合考虑，我们在主要实验中采用ROUGE指标。

We selected ROUGE as our primary text evaluation metric due to its widespread adoption and computational efficiency. However, considering that stronger text evaluation metrics are available, such as CLAIR \cite{chan2023clair} and SPICE \cite{anderson2016spicesemanticpropositionalimage}, we also report the performance of several mainstream models under these metrics in Table \ref{text quality metrics} for comprehensive reference. Specifically, the CLAIR metric employs GPT-4o as the evaluation model.

The CIDEr \cite{vedantam2015ciderconsensusbasedimagedescription} and SPICE metrics necessitate multiple reference sentences (i.e., multiple ground truths) to generate reliable evaluation outcomes. Given that RAG-IGBench provides only a single reference sentence, the reliability of these metrics is inherently compromised. While CLAIR demonstrates strong capability in differentiating text quality across various models, several considerations must be addressed: GPT-4o's potential model bias, evaluation instability arising from API modifications, and the associated computational costs. In contrast, the ROUGE metric employs rule-based text quality assessment without requiring additional large language models, thereby facilitating more efficient and cost-effective evaluation. 

\begin{table}
\vspace{-0.3cm}
    \captionof{table}{Comparison of different text quality metrics.}
    \label{text quality metrics}
    \centering
    \resizebox{0.7\textwidth}{!}{
    \begin{tabular}{lccccc}
        \toprule
        \textbf{Models} & \textbf{ROUGE} & \textbf{CIDEr} & \textbf{CIDEr-D} & \textbf{SPICE} & \textbf{CLAIR} \\
        \midrule
         GPT-4o & 56.58 & 2.458 & 0.810 & 0.0998 & 88.461 \\
         Claude-3.5-sonnet & 40.82 & 1.199 & 0.026 & 0.0584 & 87.940 \\
         Gemini-1.5-pro & 45.59 & 1.827 & 0.364 & 0.0575 & 88.113 \\
         Qwen2VL-72B & 50.92 & 2.339 & 0.549 & 0.0788 & 87.841 \\
         Qwen2VL-7B & 44.70 & 1.990 & 0.307 & 0.0880 & 84.466 \\
        \bottomrule
    \end{tabular}}
\vspace{-0.3cm}
\end{table}

In summary, while ROUGE metrics exhibit certain limitations in terms of evaluation reliability, CLAIR metrics present notable concerns regarding potential model bias and substantial additional computational overhead. Following a comprehensive assessment of these methodological trade-offs, we employ ROUGE metrics as our primary evaluation framework for the experimental analysis.

\section{Limitation}
Our RAG-IGBench, while demonstrating promising capabilities in evaluating the interleaved content generated by RAG-based methodologies, faces several important limitations that warrant discussion. First, although using model-generated answers as ground truth could potentially introduce biases, we have effectively mitigated this concern by employing diverse state-of-the-art models as generating models 
complemented by expert manual refinements. Second, while the data originates exclusively from the Xiaohongshu platform, this actually enhances content consistency while still providing substantial diversity—our dataset encompasses queries across nine categories including culture, health, education, etc., as demonstrated in the figure \ref{category}. The inherent high-quality standards of Xiaohongshu content, combined with our meticulous annotation process, ensure exceptional dataset quality despite the single-platform source. Third, regarding the predominantly Chinese original corpus, our utilization of SOTA LLMs like GPT-4o for translation ensures a high-fidelity English version with minimal semantic discrepancies. Finally, although the benchmark has been carefully curated to ensure high-quality question-answer pairs and feasible evaluation metrics, this comes at the expense of scale, thereby limiting its utility for pre-training purposes.

\section{Conclusion}

In this paper, we introduced RAG-IGBench, a novel benchmark designed specifically for evaluating interleaved image-text generation. Our benchmark advances the field through 4 key contributions: (1) A novel RAG-based approach for interleaved image-text generation. (2) A systematic benchmark with a meticulously curated dataset featuring diverse multimodal content. (3) Comprehensive evaluation metrics that assess text quality, image quality, and image-text coherence. (4) Extensive experimental analysis of both open-source and proprietary MLLMs. Our experimental results not only validate the benchmark's effectiveness but also reveal important insights into the capabilities and limitations of current MLLMs in handling complex multimodal tasks. Future research directions include expanding the benchmark's scale while maintaining its high-quality standards and developing enhanced evaluation methodologies for emerging multimodal generation tasks.

% \section*{Acknowledgments}
% This research was supported by the xxx.

\bibliographystyle{plain}  %设置参考文献类型
\bibliography{ref}      %声明参考文献文件名称

\newpage
\appendix

\section*{Appendix}

\section{Data Resource and Statistics during the data construction process}\label{limitaion}
\textbf{Resource.} All data in our benchmark are sourced from Xiaohongshu, a popular Chinese social media platform. We have conducted a rigorous content check to ensure that all data does not involve personal privacy, patents, or intellectual property issues. The personnel involved in manual annotation are professional annotators who have undergone specialized training. Importantly, all annotators are active users of the Xiaohongshu platform, making them highly familiar with the platform's image-text content and user preferences.

\begin{wraptable}{r}{8.5cm}
\vspace{-0.3cm}
    \captionof{table}{Data statistics during the data construction process.}
    \label{stat 2}
    \centering
    \resizebox{0.6\textwidth}{!}{
    \begin{tabular}{lccc}
        \toprule
        \textbf{Statistics Of Raw Cases} & \textbf{Number} & \textbf{Percentage} \\
        \midrule
        Total raw cases & 18634 & 100\% \\
        Generated by GPT-4o & 9054 & 48.58\% \\
        Generated by Claude-3.5-sonnet & 5550 & 29.78\% \\
        Generated by Gemini-1.5-pro & 4032 & 21.63\% \\
        Query w. visual demands & 17768 & 95.35\% \\
        Answer w. images & 18321 & 98.32\% \\
        High text quality & 11536 & 61.91\% \\
        High image quality & 10400 & 55.81\% \\
        High image-text consistency & 7277 & 39.05\% \\
        High overall quality & 5334 & 28.67\% \\
        Refined cases & 723 & 3.88\% \\
        \midrule
        Final dataset & 6057 & 32.51\% \\
        \bottomrule
    \end{tabular}}

\end{wraptable}

\textbf{Statistics.} We provide detailed data statistics during the data construction process in Table \ref{stat 2}. Since we pre-filtered queries without visual requirements, most queries in the human annotation exhibit visual needs, and most model-generated answers contain images. During the annotation process, we categorized text quality, image quality, image-text consistency, and overall quality into three levels: 0, 1, and 2. Note that image quality and image-text consistency are scored individually for each image. "High text quality" and "High overall quality" refer to cases with a score of 2, while "High image quality" and "High image-text consistency" indicate an average score above 1.5 per image. "Refined cases" refers to instances where we rewrote cases that had only one low-scoring dimension while all other dimensions scored high, thereby obtaining additional ground-truth data. Finally, regarding the source distribution of ground truth, we employed three state-of-the-art closed-source MLLMs (GPT-4o, Claude-3.5-Sonnet, and Gemini-1.5-Pro) for raw data generation, thus mitigating bias issues from relying on a single model and achieving more robust and fair evaluation.

\begin{figure}[h]
\centering
\includegraphics[width=14cm]{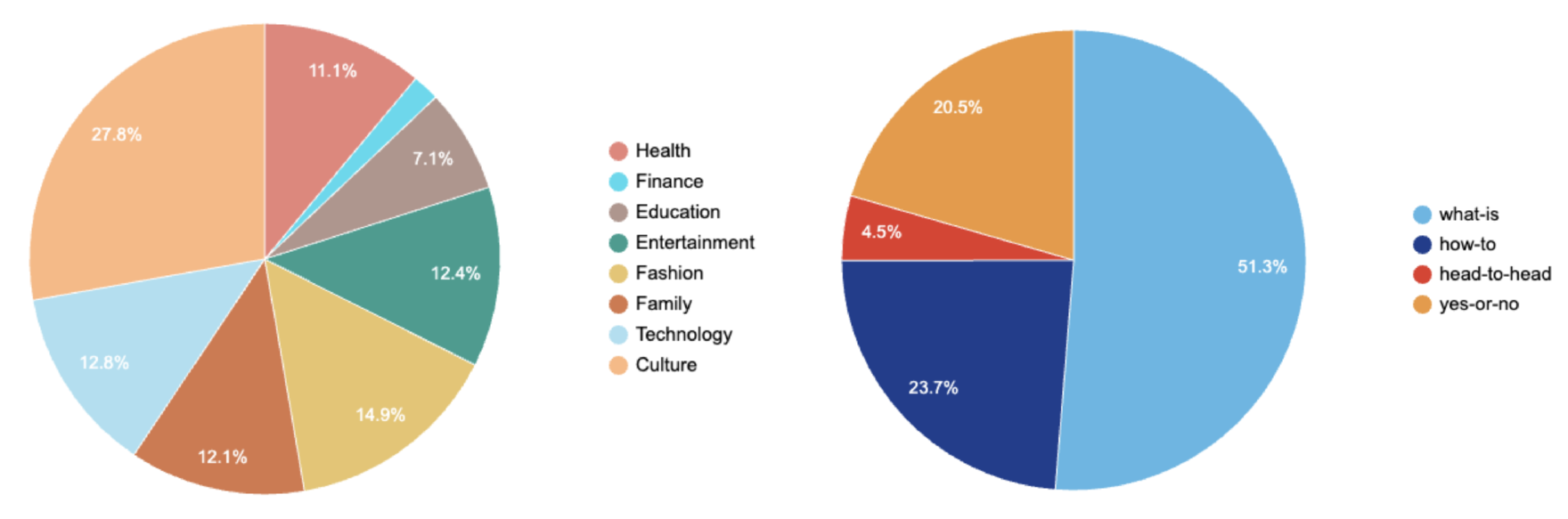}
\caption{The distribution of distinct types of queries in RAG-IGBench.}
\label{category}
\end{figure}

\section{Detailed Annotation Process}\label{detailed_anno}
Section \ref{3.3} has already provided an overview of the annotation process. The following section offers supplementary details regarding the first phase: raw answer generation.

In the first stage, we guide MLLMs (including GPT4o\cite{hurst2024gpt}, Claude-3.5-Sonnet\cite{TheC3}, et al.) with meticulously crafted instructions to generate the raw question-answer data. This stage takes a query, retrieved documents, and relevant images as input into an MLLM, resulting in a Markdown-formatted answer that includes image indices. To improve answer correctness and ensure consistency between text and images, we took inspiration from the Chain of Thought (CoT) methodology \cite{wei2023chainofthoughtpromptingelicitsreasoning}, which begins by instructing the model to analyze the input query, categorizing it into one of four types: 
\begin{itemize}[itemsep=0pt,topsep=0pt,parsep=0pt]
    \item The "what-is" query requires a direct response, ideally supplemented with an image to provide a clearer and more vivid understanding of the entity. For example, "What are the iconic landmarks in Zhengzhou?".
    \item The "how-to" query calls for a detailed, step-by-step action plan, complete with illustrations for each step to effectively guide the user through the process. For example, "How to tell if a carnation is a cold storage flower?".
    \item The "yes-or-no" query requires a straightforward affirmative or negative answer, complemented with images to enhance credibility while explaining the reason. For example, "Is \textit{Gonjiam: Haunted Asylum} worth watching?".
    \item The "head-to-head" query involves comparisons, and it is most effective to present these with visual side-by-side images for immediate clarity. For example, "What is the difference between Calla Lily and Taro?".
\end{itemize}
The main reason for categorizing queries is that different types of queries have varying requirements for visual information. After query analysis, the model engages in thoughtful reasoning to plan the logic and structure of the answer. Figure \ref{category} shows the distribution of four query categories. To further facilitate model comprehension, we incorporate an example in the instructions, in a manner analogous to the ICL approach \cite{dong2024surveyincontextlearning}. For specific prompt details, please refer to \ref{prompt}.

\section{Dataset and Code Release}\label{data and code}
Following NeurIPS Dataset and Benchmark Track guidelines, we publicly release the RAG-IGBench dataset on HuggingFace at this link: \textcolor{blue}{\url{https://huggingface.co/datasets/Muyi13/RAG-IGBench}}. RAG-IGBench is available under the Creative Commons Attribution License (CC BY 4.0). 

We include example code and instructions for (1) inference using closed- and open-source models, (2) automatic evaluation of generated answers, and (3) examining and analyzing the evaluated results. Our evaluation and metric computation scripts, along with responses from all closed- and open-source models, are accessible in the GitHub repository \textcolor{blue}{\url{https://github.com/zry13/RAG-IGBench}} under the Apache 2.0 license. We include the dataset card and README for the resources

\section{Additional Experiment of Reasoning MLLMs}
Generating content that seamlessly integrates textual and visual modalities poses significant challenges. Recently, research into the reasoning capabilities of MLLMs has intensified rapidly, making it particularly meaningful to examine their potential to enhance performance in such tasks. Below, we present additional experiments conducted on four different multimodal reasoning models. Specifically, VL-Rethinker-7B \cite{wang2025vl} and Vision-R1-7B \cite{huang2025vision} are open-source models obtained by fine-tuning QwenVL2.5-7B with GRPO and other techniques, whereas Claude3.7 (claude-3-7-sonnet-20250219) and OpenAI o1 (o1-2024-12-17) are closed-source models. For better comparison, we also include the results of QwenVL2.5-7b, GPT-4o, and Claude3.5-sonnet from the main experiment, and the reasoning MLLMs are highlighted in bold in the table.

\begin{table}[htbp]
\centering
\caption{Experiments of Reasoning MLLMs on RAG-IGBench. $\uparrow$ indicates that higher values are better.}
\label{tab:model_comparison}
\resizebox{\textwidth}{!}{
\begin{tabular}{lcccccc}
\toprule
\textbf{Models} & \textbf{Rouge-1$\uparrow$} & \textbf{Edit Distance$\uparrow$} & \textbf{Kendall Score$\uparrow$} & \textbf{Alignment Score$\uparrow$} & \textbf{CLIP Score$\uparrow$} & \textbf{Mean$\uparrow$} \\
\midrule
Qwen2.5VL-7B & 44.79 & 24.56 & 19.75 & 19.87 & 25.45 & 26.88 \\
VL-Rethinker-7B \cite{wang2025vl} & 46.70 & 27.42 & 23.24 & 21.54 & 28.99 & 29.58 \\
Vision-R1-7B \cite{huang2025vision} & 41.11 & 11.54 &  9.26 &  9.00 & 12.28 & 16.64 \\
Claude3.5-sonnet & 35.98 & 29.98 & 21.91 & 30.68 & 35.56 & 30.81 \\
Claude3.7-sonnet-thinking & 38.19 & 36.02 & 25.34 & 45.43 & 39.19 & 36.83 \\
GPT4o & 57.42 & 51.28 & 46.50 & 38.81 & 36.04 & 46.01 \\
OpenAI o1 & 38.07 & 39.16 & 35.25 & 35.22 & 40.48 & 37.64 \\
\bottomrule
\end{tabular}}

\end{table}

Both VL-Rethinker-7B and Vision-R1-7B are built upon the Qwen2.5VL-7B foundation model and trained using GRPO. However, VL-Rethinker-7B achieves a 2.7\% improvement in overall performance, while Vision-R1-7B shows a 10.24\% decline in comprehensive performance. We attribute this disparity primarily to their distinct training paradigms: Vision-R1-7B is predominantly optimized for mathematical reasoning tasks, while VL-Rethinker-7B incorporates a broader spectrum of real-world reasoning scenarios. This limited scope is the primary reason for Vision-R1-7B's inferior performance.

Regarding proprietary reasoning models, Claude-3.7-Sonnet-thinking with reasoning capabilities significantly outperforms the standard Claude-3.5-Sonnet on RAG-IGBench, demonstrating that enhanced reasoning abilities can improve performance on the RAG-IG task. Although the o1 model performs considerably worse than GPT-4o, we believe this is because o1's multimodal understanding capabilities are inherently weaker than those of GPT-4o.

\section{GRPO Experiment Results Based on Our Metrics}
Recent developments in the DeepSeek-R1 \cite{deepseekai2025deepseekr1incentivizingreasoningcapability} series of work introduced Group Relative Policy Optimization (GRPO), a novel training paradigm that employs rule-based reward functions to enhance the reasoning capabilities of LLMs significantly. Subsequently, numerous studies \cite{shen2025vlmr1stablegeneralizabler1style, chen2025r1v, huang2025visionr1incentivizingreasoningcapability} have attempted to extend GRPO to MLLMs. However, most of these efforts have been limited to tasks with easily verifiable answers, such as Math, OVD, and REC. The reason is that open-domain question answering presents challenges for GRPO implementation due to the difficulty in efficiently evaluating responses through rule-based approaches.

Our RAG-IGBench addresses this limitation through the innovative metrics we proposed, which effectively solve the evaluation challenge. Leveraging these newly developed metrics, we have designed novel reward functions capable of scoring generated multimodal responses, thereby enabling GRPO experiments in open-domain multimodal contexts. 

Table \ref{grpo exp} presents our preliminary experimental results. We conducted Supervised Fine-Tuning (SFT) and Group Relative Policy Optimization (GRPO) experiments on both the 3B and 7B versions of the Qwen2.5VL-Instruct model. We compared the different improvements brought by SFT and GRPO, and further performed GRPO training on the SFT checkpoint to demonstrate the compatibility between these two approaches. For SFT training, we set the initial learning rate to 1e-5 with a warmup ratio of 0.05, and trained for a total of 3 epochs. For GRPO training, we set the generation number to 8 and trained for 5,000 steps. We designed the reward model and training scripts based on the VLM-R1 \cite{shen2025vlmr1stablegeneralizabler1style} codebase, and we acknowledge the open-source work of \cite{shen2025vlmr1stablegeneralizabler1style}. The training scripts for this experiment have also been made publicly available.

\begin{table}[h]
    \centering
    \resizebox{\textwidth}{!}{
    \begin{tabular}{lccccc|>{\columncolor{gray!20}}c}
        \toprule
        \textbf{Models} & \textbf{Text Quality} & \multicolumn{2}{c}{\textbf{Image Quality}}  & \multicolumn{2}{c}{\textbf{Image-Text Consistency}}  & \textbf{Mean}\( \uparrow \) \\
        \midrule
         & \textbf{Rouge-1}\( \uparrow \) & \textbf{\makecell[c]{Edit \\ Distance}}\( \uparrow \) & \textbf{\makecell[c]{Kendall \\ Score}}\( \uparrow \) & \textbf{\makecell[c]{Alignment \\ Score}}\( \uparrow \) & \textbf{\makecell[c]{CLIP \\ Score}}\( \uparrow \) & \\
        \midrule
        Qwen2.5VL-3B-Instruct & 41.49 & 17.71 & 10.44 & 16.78 & 19.80 & 16.30 \\
        \hspace{2em}+ SFT & 50.81 & 32.99 & 22.81 & 35.31 & 39.15 & 33.11 \\
        \hspace{2em}+ GRPO & 42.01 & 31.66 & 24.23 & 30.52 & 41.20 & 31.63 \\
        \hspace{2em}+ SFT and GRPO & 50.95 & 31.65 & 23.36 & 35.74 & 40.10 & 33.31 \\
        \midrule
        Qwen2.5VL-7B-Instruct & 45.82 & 24.41 & 17.07 & 22.41 & 28.18 & 22.89 \\
        \hspace{2em}+ SFT & 52.91 & 34.35 & 24.30 & 35.47 & 39.25 & 33.77 \\
        \hspace{2em}+ GRPO & 45.43 & 31.19 & 25.22 & 33.41 & 41.77 & 33.00 \\
        \hspace{2em}+ SFT and GRPO & 52.35 & 34.63 & 24.69 & 35.38 & 38.95 & 33.81 \\
        \bottomrule
    \end{tabular}}
    \caption{The experiment result of GRPO training based on our innovative metrics.}
    \label{grpo exp}
\end{table}

The experimental results demonstrate that GRPO training yields improvements in the dimensions of image quality and image-text consistency, while showing minimal changes in text quality. Compared to SFT, GRPO achieves greater enhancements in Kendall score and CLIP score, though the improvement in edit distance is relatively modest. Furthermore, the results of the combined SFT+GRPO training confirm the compatibility of these two training approaches. We believe this is because GRPO enhances the model's ability to understand multiple images, whereas SFT primarily improves the model's textual capabilities.

\section{Prompt Templates}\label{prompt}
Here, we present the prompt template used for generating interleaved image-text content. We employs Jinja2 as a templating system, where during inference, the prompt is populated with specific queries and relevant image-text content to form complete instructions.

\lstset{
    basicstyle=\ttfamily\small, % 设置代码字体和大小
    backgroundcolor=\color{gray!20}, % 设置背景颜色为灰色（20% 灰度）
    rulesepcolor=\color{gray}, % 边框颜色
    breaklines=true, % 启用自动换行
    captionpos=b, % 标题位置在代码块下方
    numbers=left, % 行号在左侧
    numberstyle=\tiny\color{gray}, % 行号样式
    keywordstyle=\color{blue}, % 关键字颜色
    commentstyle=\color{green}, % 注释颜色
    stringstyle=\color{red} % 字符串颜色
}
\begin{lstlisting}
You are an answer aggregation system that generates informative answers based on the user's query and relevant documents.

User queries can be categorized into the following types:
1) what-is: definition type queries, for which the answer should be direct and include entity images when possible to make the response more illustrative;
2) how-to: procedure type queries, for which the answer should outline a process with steps preferably accompanied by corresponding images;
3) yes-or-no: verification type queries, for which the answer should be direct;
4) head-to-head: comparison type queries, for which the answer should be presented in a table format.

Below you will receive a specific user query, relevant numbered documents (e.g., DOC#1), and corresponding numbered images (e.g., IMG#1). Please determine the query category, provide an ideal answer, include appropriate reasoning, and insert suitable images to enhance your answer when necessary.

Input and output should follow the JSON format below:

Input:
{
    "query": "xxx", # User query, str
    "documents and corresponding images": [    # Retrieved relevant documents and corresponding images, List[dict]
        {
            "document": "DOC#1\nxxx",    # Relevant numbered document
            "images": [IMG#1, IMG#2, xxx]    # Corresponding numbered images list
        },
        {
            "document": "DOC#2\nxxx",
            "images": [IMG#3, IMG#4, xxx]
        }
    ]
}

Output:
{
    "reason": "xxx", # Rationale for determining query category, approach for organizing the answer, and layout planning, str
    "category": "xxx", # Query category, must be one of ["what-is", "how-to", "yes-or-no", "head-to-head"], str
    "answer": "xxx" # Ideal answer, str
}

The generated answer should meet the following requirements:
1. Use **markdown** syntax with appropriate formatting;
2. Properly cite documents as superscript references, e.g., xxx<sup>[3](DOC#3)</sup>xxx, where DOC#3 is the document number. Avoid explicitly mentioning phrases like 'data source' or 'according to the document'. Each document should be cited only once;
3. Insert key images as core content, e.g., xxx![dummy](IMG#1)xxx, where IMG#1 represents the first image. Image placement should be appropriate, selection should be relevant, placement should be precise, multiple images should maintain consistent style, and each image should be inserted only once;
4. Formatting should prioritize rationality, aesthetics, and simplicity.

Here is a question-answering example:
{
    "query": "difference between polo and Ralph Lauren"
    Output: "The user query is asking about the difference between Polo and Ralph Lauren, which falls into the image-text category. By providing comparative logo images of the brands, we can visually demonstrate their differences while complementing with brief textual explanations to better address the user's question.",
    "category": "image-text",
    "answer": "The differences between Polo and Ralph Lauren are mainly reflected in their brands and logo designs. Ralph Lauren is an American luxury brand, and its men's clothing line is called Polo Ralph Lauren, with a logo featuring a single rider on a single horse, holding a mallet in the right hand and leaning toward the left of the image<sup>[1](DOC#1)</sup>. Polo Sport is a domestic copycat brand, also with a single rider on a single horse logo, but holding a mallet in the left hand and leaning toward the right of the image<sup>[2](DOC#2)</sup>. Additionally, there's U.S. POLO ASSN., whose logo features two riders on two horses, with one rider positioned in front of the other<sup>[3](DOC#3)</sup>. The image below shows the differences between these brand logos: ![logo comparison](IMG#1)"
}

Here is the given input:
{
    "query": {{query}},
    "documents and corresponding images": [
        {% for note in notes %}
        {
            "document": "{{note['doc']}}",
            "images": {{note['img']}}
        }
        {% endfor %}
    ]
}

Following the question-answering example, and based on the provided documents and image information, please provide your output. Remember to appropriately cite documents in your answer, insert relevant images in suitable positions when necessary, and ensure your output follows the JSON format.
\end{lstlisting}

\section{Qualitative Example}

\subsection{Comparative generation results across different MLLMs}
\begin{figure}[h]
    \centering
    \includegraphics[width=\textwidth]{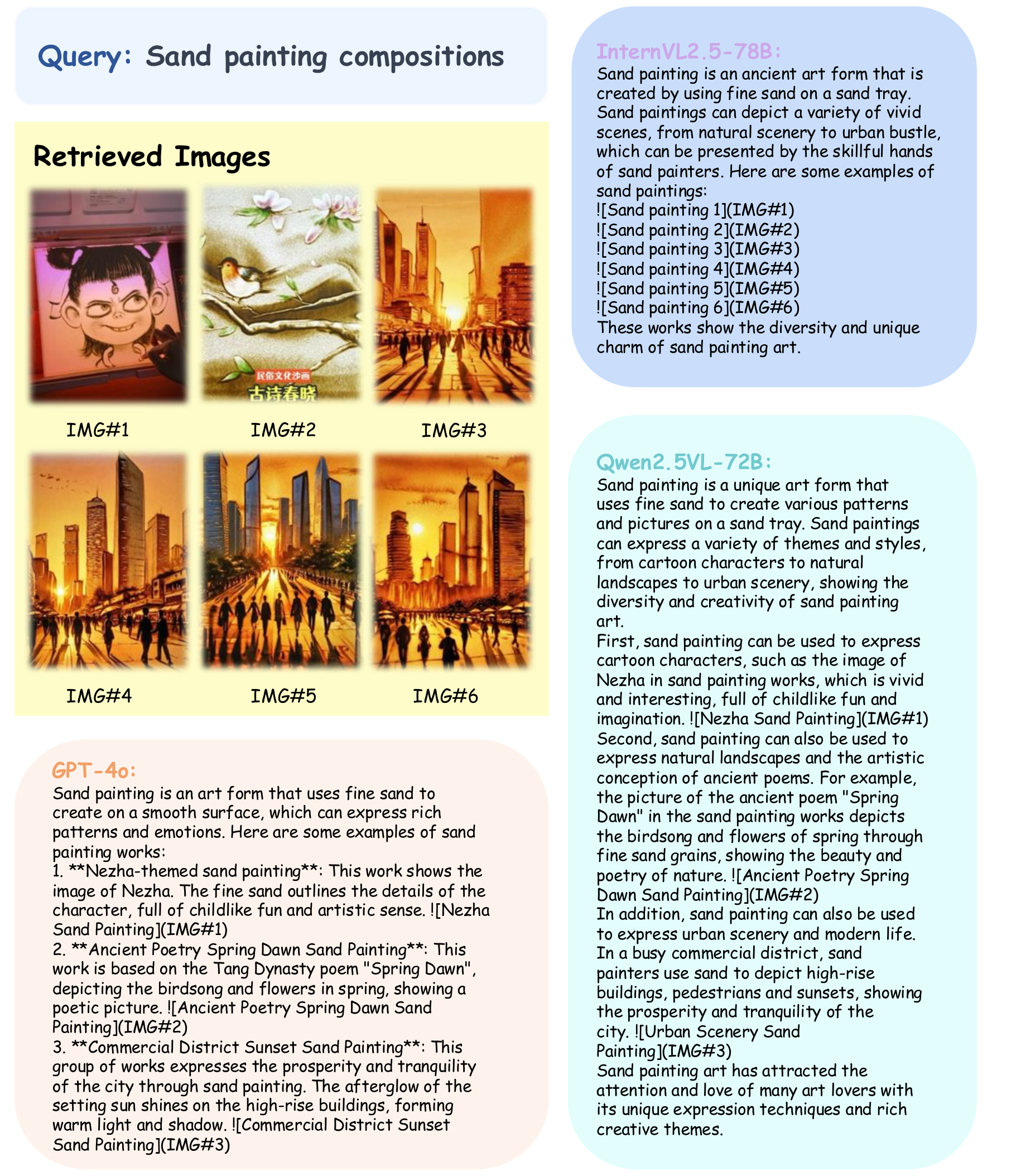}
    \caption{Comparison of results generated by different models}
    \label{Comparative}
\end{figure}

As illustrated in the figure \ref{Comparative}, we show the outputs generated by distinct multimodal large language models responding to identical queries under our instructions. While all models successfully produced syntactically valid markdown-formatted responses, they exhibit notable variations in their presentation of image-text integration and visual-textual alignment.

\begin{figure}[t]
    \centering
    \includegraphics[width=\textwidth]{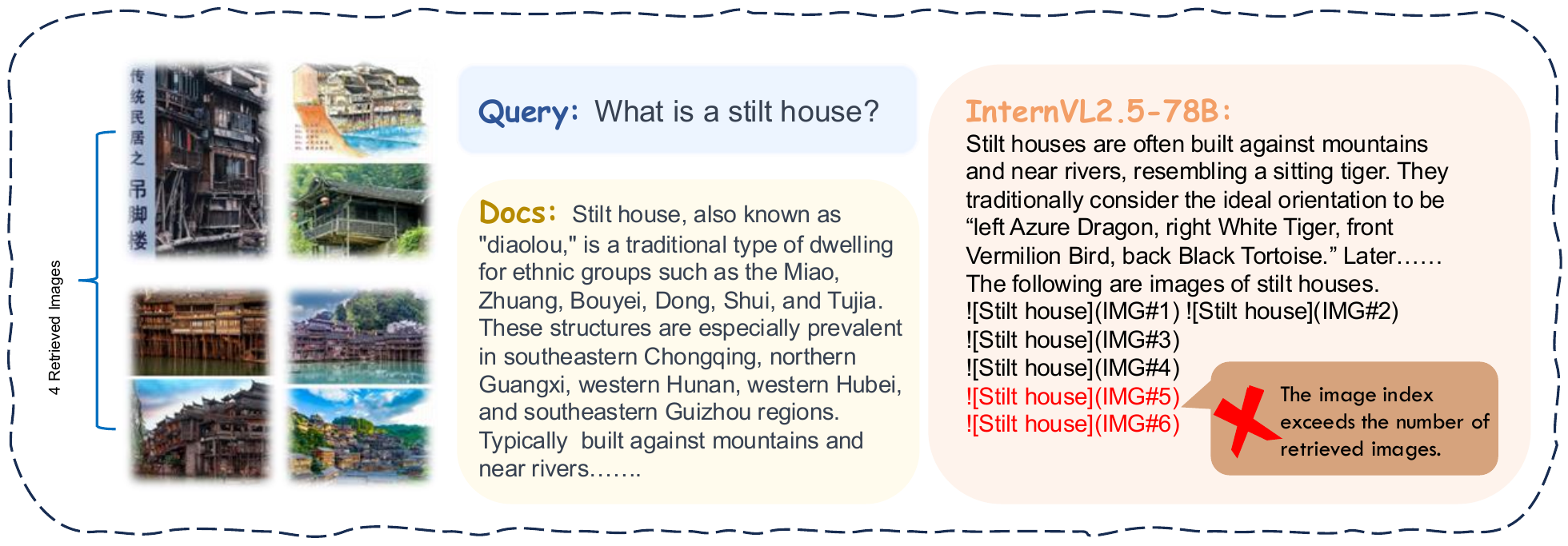}
    \caption{Examples of model failures.}
    \label{error case}
\end{figure}

\subsection{Failure examples}

The figure \ref{error case} illustrates an example of model error, demonstrating instances where the generated responses reference image indices that exceed the number of retrieved images. As analyzed in section \ref{4.3}, this represents a form of hallucination phenomenon. These findings highlight that addressing hallucinations in multimodal large language models constitutes a critical research direction. The observed inconsistencies between textual outputs and visual inputs underscore the challenges in achieving reliable cross-modal reasoning, suggesting that developing robust mechanisms to mitigate such hallucinations remains an important frontier in advancing multimodal AI systems.

\newpage
\section*{NeurIPS Paper Checklist}

\begin{enumerate}

\item {\bf Claims}
    \item[] Question: Do the main claims made in the abstract and introduction accurately reflect the paper's contributions and scope?
    \item[] Answer: \answerYes{} % Replace by \answerYes{}, \answerNo{}, or \answerNA{}.
    \item[] Justification: The abstract and introduction clearly reflect the main contributions and scope of this paper, elucidating a novel approach and an innovative evaluation framework proposed in the field of multimodal generation.

\item {\bf Limitations}
    \item[] Question: Does the paper discuss the limitations of the work performed by the authors?
    \item[] Answer: \answerYes{} % Replace by \answerYes{}, \answerNo{}, or \answerNA{}.
    \item[] Justification: See appendix \ref{limitaion}.

\item {\bf Theory assumptions and proofs}
    \item[] Question: For each theoretical result, does the paper provide the full set of assumptions and a complete (and correct) proof?
    \item[] Answer: \answerNA{} % Replace by \answerYes{}, \answerNo{}, or \answerNA{}.
    \item[] Justification: The paper does not include theoretical results. 

\item {\bf Experimental result reproducibility}
    \item[] Question: Does the paper fully disclose all the information needed to reproduce the main experimental results of the paper to the extent that it affects the main claims and/or conclusions of the paper (regardless of whether the code and data are provided or not)?
    \item[] Answer: \answerYes{} % Replace by \answerYes{}, \answerNo{}, or \answerNA{}.
    \item[] Justification: This paper provides open access to the source code and datasets, along with detailed documentation of the experimental procedures

\item {\bf Open access to data and code}
    \item[] Question: Does the paper provide open access to the data and code, with sufficient instructions to faithfully reproduce the main experimental results, as described in supplemental material?
    \item[] Answer: \answerYes{} % Replace by \answerYes{}, \answerNo{}, or \answerNA{}.
    \item[] Justification: See appendix \ref{data and code}.

\item {\bf Experimental setting/details}
    \item[] Question: Does the paper specify all the training and test details (e.g., data splits, hyperparameters, how they were chosen, type of optimizer, etc.) necessary to understand the results?
    \item[] Answer: \answerYes{} % Replace by \answerYes{}, \answerNo{}, or \answerNA{}.
    \item[] Justification: The paper outlines the key configurations and methodological settings implemented in \ref{exp}. More granular information is provided in the publicly released codebase.

\item {\bf Experiment statistical significance}
    \item[] Question: Does the paper report error bars suitably and correctly defined or other appropriate information about the statistical significance of the experiments?
    \item[] Answer: \answerYes{} % Replace by \answerYes{}, \answerNo{}, or \answerNA{}.
    \item[] Justification: Though we were unable to conduct extensive repeated evaluations due to computational resource limitations, we provide the data statistics as shown in the table \ref{data stas}.

\item {\bf Experiments compute resources}
    \item[] Question: For each experiment, does the paper provide sufficient information on the computer resources (type of compute workers, memory, time of execution) needed to reproduce the experiments?
    \item[] Answer: \answerYes{} % Replace by \answerYes{}, \answerNo{}, or \answerNA{}.
    \item[] Justification: See section \ref{exp}.

\item {\bf Code of ethics}
    \item[] Question: Does the research conducted in the paper conform, in every respect, with the NeurIPS Code of Ethics \url{https://neurips.cc/public/EthicsGuidelines}?
    \item[] Answer: \answerYes{} % Replace by \answerYes{}, \answerNo{}, or \answerNA{}.
    \item[] Justification: Our research adheres to the NeurIPS Code of Ethics.

\item {\bf Broader impacts}
    \item[] Question: Does the paper discuss both potential positive societal impacts and negative societal impacts of the work performed?
    \item[] Answer: \answerYes{} % Replace by \answerYes{}, \answerNo{}, or \answerNA{}.
    \item[] Justification: See appendix \ref{limitaion}.

\item {\bf Safeguards}
    \item[] Question: Does the paper describe safeguards that have been put in place for responsible release of data or models that have a high risk for misuse (e.g., pretrained language models, image generators, or scraped datasets)?
    \item[] Answer: \answerNA{} % Replace by \answerYes{}, \answerNo{}, or \answerNA{}.

\item {\bf Licenses for existing assets}
    \item[] Question: Are the creators or original owners of assets (e.g., code, data, models), used in the paper, properly credited and are the license and terms of use explicitly mentioned and properly respected?
    \item[] Answer: \answerYes{} % Replace by \answerYes{}, \answerNo{}, or \answerNA{}.
    \item[] Justification: All codes, datasets, and models used in the paper have been properly cited with their original sources.

\item {\bf New assets}
    \item[] Question: Are new assets introduced in the paper well documented and is the documentation provided alongside the assets?
    \item[] Answer: \answerYes{} % Replace by \answerYes{}, \answerNo{}, or \answerNA{}.
    \item[] Justification: The new assets introduced in this paper are accompanied by documentation, which is provided alongside the assets and includes detailed instructions for dataset evaluation and usage.

\item {\bf Crowdsourcing and research with human subjects}
    \item[] Question: For crowdsourcing experiments and research with human subjects, does the paper include the full text of instructions given to participants and screenshots, if applicable, as well as details about compensation (if any)? 
    \item[] Answer: \answerNA{} % Replace by \answerYes{}, \answerNo{}, or \answerNA{}.
    \item[] Justification: This paper does not involve crowdsourcing or research with human subjects.

\item {\bf Institutional review board (IRB) approvals or equivalent for research with human subjects}
    \item[] Question: Does the paper describe potential risks incurred by study participants, whether such risks were disclosed to the subjects, and whether Institutional Review Board (IRB) approvals (or an equivalent approval/review based on the requirements of your country or institution) were obtained?
    \item[] Answer: \answerNA{} % Replace by \answerYes{}, \answerNo{}, or \answerNA{}.
    \item[] Justification: This paper does not involve crowdsourcing or research with human subjects.

\item {\bf Declaration of LLM usage}
    \item[] Question: Does the paper describe the usage of LLMs if it is an important, original, or non-standard component of the core methods in this research? Note that if the LLM is used only for writing, editing, or formatting purposes and does not impact the core methodology, scientific rigorousness, or originality of the research, declaration is not required.
    %this research? 
    \item[] Answer: \answerNA{} % Replace by \answerYes{}, \answerNo{}, or \answerNA{}.
    \item[] Justification: The development of the core methods in this research does not involve LLMs as any important, original, or non-standard components.

\end{enumerate}

\end{document}